\documentclass[10pt,aps,prl,twocolumn,superscriptaddress]{revtex4-1}

\usepackage{amsmath}
\usepackage{textcomp}
\usepackage{amssymb} 
\usepackage{graphicx}
 
\usepackage{tabularx}
\usepackage{placeins}
\usepackage{natbib}
\usepackage[normalem]{ulem}
\usepackage[usenames]{color}

\usepackage[colorlinks=true,linkcolor=blue,citecolor=blue]{hyperref}%

\begin{document}

\preprint{APS/123-QED}

\title{Effects of local incompressibility on the rheology of composite biopolymer networks}

\author{Anupama Gannavarapu}
\affiliation{Department of Chemical \& Biomolecular Engineering, Rice University, Houston, Texas 77005, USA}
\affiliation{Center for Theoretical \& Biophysics, Rice University, Houston, Texas 77005, USA}
\author{Sadjad Arzash}
\affiliation{Department of Physics, Syracuse University, Syracuse, New York 13244, USA}
\affiliation{Department of Physics \& Astronomy, University of Pennsylvania, Philadelphia, Pennsylvania 16802, USA}
\author{Iain Muntz}
\affiliation{Department of Bionanoscience, Kavli Institute of Nanoscience, Delft University of Technology, Van der Maasweg 9, 2629 HZ Delft, Netherlands}
\author{Jordan L. Shivers}
\affiliation{Department of Chemistry, University of Chicago, Chicago, Illinois 60637, USA}
\affiliation{James Franck Institute, University of Chicago, Chicago, Illinois 60637, USA}
\author{Anna-Maria Klianeva}
\affiliation{Department of Bionanoscience, Kavli Institute of Nanoscience, Delft University of Technology, Van der Maasweg 9, 2629 HZ Delft, Netherlands}
\author{Gijsje H. Koenderink}
\affiliation{Department of Bionanoscience, Kavli Institute of Nanoscience, Delft University of Technology, Van der Maasweg 9, 2629 HZ Delft, Netherlands}
\author{Fred C. MacKintosh}
\affiliation{Department of Chemical \& Biomolecular Engineering, Rice University, Houston, Texas 77005, USA}
\affiliation{Center for Theoretical \& Biophysics, Rice University, Houston, Texas 77005, USA}
\affiliation{Departments of Chemistry, Rice University, Houston, Texas 77005, USA}
\affiliation{Departments of Physics \& Astronomy, Rice University, Houston, Texas 77005, USA}

\date{\today}

\begin{abstract}
Fibrous networks such as collagen are common in biological systems. Recent theoretical and experimental efforts have shed light on the mechanics of single component networks. Most real biopolymer networks, however, are composites made of elements with different rigidity. For instance, the extracellular matrix in mammalian tissues consists of stiff collagen fibers in a background matrix of flexible polymers such as hyaluronic acid (HA). The interplay between different biopolymer components in such composite networks remains unclear. In this work, we use 2D coarse-grained models to study the nonlinear strain-stiffening behavior of composites. We introduce a local volume constraint to model the incompressibility of HA. We also perform rheology experiments on composites of collagen with HA. We demonstrate both theoretically and experimentally that the linear shear modulus of composite networks can be increased by approximately an order of magnitude above the corresponding moduli of the pure components. Our model shows that this synergistic effect can be understood in terms of the local incompressibility of HA, which acts to suppress density fluctuations of the collagen matrix with which it is entangled.
\end{abstract}

\maketitle


\section{\label{sec:introduction} Introduction}

The mechanical stability of cells and tissues depends on complex interconnected biopolymer networks such as the cytoskeleton inside cells and the extracellular matrix outside cells. These networks are made of diverse structural components that work together to support physiological tasks such as cellular rearrangements, tissue growth and signaling, thereby translating to a diversity in functions and properties \cite{necas2008hyaluronic, fletcher2010cell}. Although the properties of the individual components of these composite materials vary, they complement one another for overall enhanced mechanical properties. In extracellular matrices, stiff collagen type I fibers are often found in a softer matrix of flexible polysaccharides such as hyaluronic acid (HA). As the most abundant protein in the human body, collagen forms fibrillar networks that can bear high tensile stresses \cite{jansen2018role, fratzl1998fibrillar}, while hyaluronic acid forms hydrogels that are known to resist compression, e.g., for lubrication of joints \cite{necas2008hyaluronic}.

Although biopolymers are distinct in their chemical structures, the macroscopic properties of their networks can be largely independent of the microscopic details. Recent studies have shown that the mechanics of these networks can be understood in terms of the collective behavior of simplified constituents. Specifically, coarse-grained models based on elastic springs and bending beams have been shown to lead to quantitative and predictive models of collagen network elasticity \cite{sharma2016strain, jansen2018role, Doorn2017, burla2019stress, wyse2022structural}.
An important architectural aspect of such models is the average coordination number or connectivity $\langle z \rangle$ \cite{Broedersz2014}. Collagen networks have a connectivity between $3$ and $4$ \cite{stein2008algorithm, jansen2018role, burla2020connectivity}, which places them well below the isostatic threshold for mechanical stability of $z_c = 2d$, where $d$ is the dimensionality, as originally identified by Maxwell \cite{maxwell_1870, calladine1978buckminster}. Thus, such networks are \emph{sub-isostatic} and their linear stability at small strain must depend on more than spring-like energies alone \cite{wyart2008elasticity, Broedersz2011}. In the case of collagen type I networks, the linear elasticity can be understood to be due to the bending resistance of the constituent fibers, while this bending response can transition to a stretching response of fibers at large enough strain \cite{satcher1996theoretical, kroy1996force, onck2005alternative, jansen2018role}. Moreover, this transition has recently been shown to be a second-order phase transition with rich critical behavior \cite{sharma2016strain}.

In the linear elastic regime governed by bending of fibers, the network strain must also be nonaffine, characterized by significant local nonuniformity in the strain field and a shear modulus far below that of a purely stretching response \cite{kroy1996force, head2003deformation, wilhelm2003elasticity, onck2005alternative, broedersz2012filament, chen2023nonaffine}. 
This soft bending-dominated regime can be understood to be due to the nonaffine strain. Thus, suppression of nonaffinity should generally result in network stiffening. 
Here, we study the effect of local incompressibility on the mechanics of fiber networks. 
In general, the strain field can be decomposed into volume-preserving (shear-like) and compressive or dilational components. Under applied bulk strain, only nonaffine deformations can change the local volume or density. The relative importance, however, of the nonaffine dilational component for the elasticity of networks such as collagen is not known. Given the important role of HA that is known to affect tissue-level compressibility of extracellular matrices, it seems likely that it may reduce nonaffine strain at the local level and hence affect extracellular matrix mechanics. 

Here, we develop a computational model to study the role of local incompressibility on network mechanics and we find that incompressibility alone can lead to approximately 10-fold stiffening of fibrous networks in the linear elastic regime. We also demonstrate experimentally a similar increase of the linear shear modulus of collagen networks upon the addition of HA. 
Interestingly, in contrast with prior computational models that have shown similar stiffening effects of additional elastic interactions such as bending resistance or additional Hookean springs, we show that the effect of local incompressibility is limited. Specifically, even in the limit of strong elastic suppression of local compression, the linear shear modulus remains below that for a purely stretching response of the network. This is consistent with the presence of both volume-preserving and dilational components of the nonaffine response of bending dominated fiber networks such as collagen in the linear regime. 
Our computational model also constitutes a computationally efficient model for creating elastic networks with desired bulk and shear moduli. This is in contrast to the common practice of modeling networks with Hookean interactions, which necessarily lead to comparable bulk and shear moduli for regular or disordered structures, whereas most real materials have disparate bulk and shear moduli. 
Furthermore, each of the components bring competing energies to the table including fiber stretching, bending, matrix stretching and the resistance to compressive stresses leading to a unique and non-trivial mechanical behavior. Hence, we develop a coarse-grained composite model that explains the surprising mechanical properties of composite systems under shear, often composed of soft and rigid elements.

\section{\label{sec:overview} Two-fluid model with incompressible matrix}

Collagen networks are embedded as a stiff macromolecular fiber matrix in an interpenetrating fluid with very different mechanical properties. In the case of a surrounding Newtonian liquid medium, the network response can differ substantially on different length- or timescales. 
On short timescales or on large length scales, the fluid viscosity prevents significant relative motion of network and fluid, rendering the composite system effectively incompressible along with the liquid. The Poisson ratio $\nu=1/2$ in this case.
By contrast, at longer times, a decoupling from the fluid can lead to compression and significantly smaller Poisson ratio for the network. 
In fact, for such a two-fluid system, there is no well-defined Poisson ratio \cite{brochard1977dynamical, milner1993dynamical, schnurr1997macromolecules, levine2000one, mackintosh2008nonequilibrium}.
The poroelastic timescale $\tau$ for this decoupling is expected to depend on the fluid viscosity $\eta$ and the network shear modulus $G$ and pore size $\xi$, according to $\tau\sim\eta r^2/(G\xi^2)$, where $r$ is the characteristic length-scale over which fluid transport occurs \cite{henriporosity, Vahabi2018}.

Hyaluronic acid can itself form a relatively flexible polymer hydrogel with a pore size approximately two orders less than that of collagen \cite{burla2020particle}. Pure HA networks are well-approximated as linear (visco)elastic up to strains larger than for the linear regime of collagen networks \cite{Gibbs1968, lelu2007characterization, Zhang2013, cowman2015content}. 
For collagen networks embedded in HA matrices, in addition to a continuum viscous-like coupling, one must account for a much stronger topological entanglement of collagen and HA. For high molecular weight HA, this topological entanglement can be expected to dramatically suppress relative motion of the network and embedding fluid. 
We approximate the HA gel as an elastic matrix with shear modulus $G_m$ and much larger Lam\'{e} coefficient $\lambda_m$ on relevant timescales and length-scales larger than the collagen pore size $\xi$, which is much larger than that of HA. We thus modify the prior non-inertial two-fluid model \cite{mackintosh2008nonequilibrium} equations of motion as
\begin{equation}
0 = G_m \nabla^2 \vec{u_m} + (G_m+\lambda_m) \vec{\nabla} \cdot (\vec{\nabla} \cdot \vec{u_m})  - \Gamma (\vec{u_m}-\vec{u_f})
   \label{eq:fluid}
\end{equation}
and
\begin{equation}
0 = G_f \nabla^2 \vec{u_f} + (G_f+\lambda_f) \vec{\nabla} \cdot (\vec{\nabla} \cdot \vec{u_f})  + \Gamma (\vec{u_m}-\vec{u_f}),
 \label{eq:network}
\end{equation}
where $u_m$ refers to the matrix displacement field and the second equation refers to the corresponding fiber terms. 
Here, $\Gamma$ represents the coupling between the collagen fiber network and the HA matrix. In principle, this coupling is both viscoelastic and topological in nature, but we assume it to be strong and dominated by topological constraints. 
Physically, this constraint suppresses the flow of the HA matrix from one unit cell of the fiber network to another. This may potentially dramatically reduce the the nonaffine deformation of the collagen in response to shear, as we explore below.

\section{\label{sec:model} Model}

\begin{figure}[h]
\centering
\includegraphics[width=0.45\textwidth]{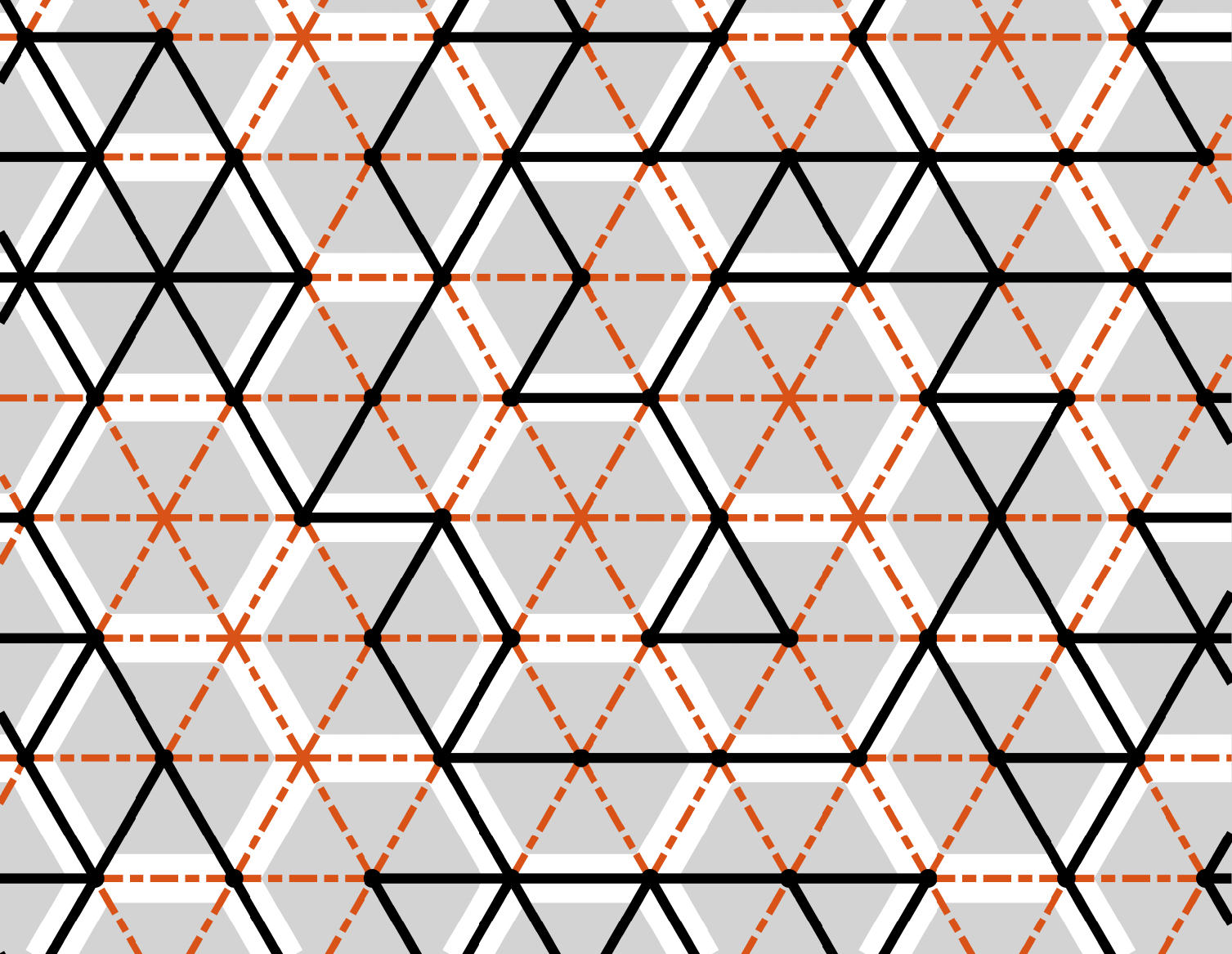}
\caption{A snapshot of the arrangement of the composite network on a triangular lattice. The dotted red lines depict an undiluted network representing the hyaluronic acid matrix.
The solid black lines illustrate a diluted triangular network ($\langle z \rangle = 3.3$) analogous to the structure of collagen and connected to the matrix at every node. A local volume constraint of area rigidity, $\beta$ is applied over sets of six triangles or a full hexagonal lattice, as shown by the grey areas in the diagram. In this system, the nodes are connected to each other with harmonic springs of moduli $\mu_m = 10^{-5}$ and $\mu_f = 1.0$ which correspond to the moduli of the matrix and fiber respectively. Additionally, the fibers have a low bending modulus of $\kappa = 10^{-4}$.}
\label{fig:network} 
\end{figure}

To investigate the properties of composite networks comprising stiff collagen fibers embedded in soft hyaluronic acid, we utilized a 2D triangular lattice as our model, following many prior studies \cite{mao2013effective, feng2016nonlinear, hatami2020mechanical, yucht2013dynamical, Broedersz2011}. Initially, we generated a periodic triangular lattice of size $W = 90$ consisting of $N = W^2$ Hookean springs to model the HA network. All the springs were directly cross-linked at the nodal points. A section of the constructed network is shown in Fig.\ \ref{fig:network}. Importantly, since this structure has a coordination number or connectivity of $z=6$, which exceeds the isostatic threshold of $z_c = 4$ in 2D \cite{calladine1978buckminster, maxwell_1870}, the response is expected to be nearly linear with both $G_m$ and $\lambda_m$ proportional to $\mu_m$. 
As has been noted, one cannot account for the expected large compression modulus (i.e., $\lambda_m\gg G_m$) with such a model. 
We can model a large compression modulus by imposing a local area constraint similar to what has been done, e.g., in prior vertex models for epithelial tissues \cite{davidson2010emergent, Merkel2018, farhadifar2007influence, bi2014energy}. 
Importantly, we add such a constraint not at the level of individual triangles but for hexagons, as shown in Fig. \ref{fig:network}, as we explain below. 

The resulting matrix hamiltonian is given by 
\begin{equation}
H_\mathrm{matrix} = \frac{\mu_m}{2} \sum_{ij}\frac{(l_{ij} - l_{ij,0})^2}{l_{ij,0}} + \frac{\beta}{2} \sum_{h}\frac{(A_{h} - A_{h,0})^2}{A^2_{h,0}}  \label{eq:H_matrix}
\end{equation}
where $l_{ij,0}=1$ represents the length of the $ij$ bond/spring in the relaxed state, $l_{ij}$ is the current bond length in the deformed state, $A_{h,0}$ is the initial relaxed area of the hexagon over which the area constraint is being applied and $A_h$ is the current area of the hexagon. Here, $\beta$ represents the Hookean-like strength of the applied local volume constraint, which can be expected to determine the bulk modulus $B$ of the resulting network. 
We confirm this below. 
As shown in Eq.\ \eqref{eq:H_matrix}, the local volume incompressibility of the matrix was modeled as a quadratic energy cost due to the change in area, which can also be understood as a penalty imposed when the network moves away from the equilibrium area (here, in 2D) at a local level. To implement this local constraint, we divided the matrix network into hexagonal unit cells and applied the penalty in energy for each of these units. Imposing an additional constraint onto hexagonal structures instead of each of the triangles that compose them avoids volumetric locking and over-constraining the system. When a structure is subjected to volumetric locking, it develops very high stresses and is devoid of all floppy modes \cite{colom2015mpm}. In other words, it would exhaust the system from all of the degrees of freedom, curtailing node movement. This would also be contradictory to our ultimate objective of constructing a composite model that exhibits distinguishable shear and bulk moduli. In addition, using a hexagonal structure on a triangular architecture provides an added advantage of preserving rotational symmetry, which is not the case with other regular polygons of comparable vertex count.

The fibers were then added to the matrix using springs of a higher stretch modulus, and a bending rigidity, $\kappa$. To avoid the effect of spanning fibers, each of the fibers were initially cut at a random bond. The fibers were then randomly diluted by cutting random bonds until the fiber network had an average sub-isostatic connectivity of $\langle z \rangle = 3.3$, which is below the Maxwell isostatic threshold of $z_c \sim 2d = 4$ (\textit{d} dimensions) in 2D. Any dangling ends (nodes with only one connection) that have no effect on the mechanics of these networks were removed. 

The matrix comprises a stretching energy and an energy resulting from local volume preservation. The fiber energy includes stretching energy of the springs and a bending energy calculated as the resistance to bending between two nearest-neighbor bonds on the same fiber. 
The Hamiltonian for the fiber networks can be written as \cite{Broedersz2014, Merkel2018},
\begin{equation}
H_\mathrm{fiber} = \frac{\mu_f}{2} \sum_{ij}\frac{(l_{ij} - l_{ij,0})^2}{l_{ij,0}} +  \frac{\kappa_f}{2} \sum_{ijk}\frac{(\theta_{ijk} - \theta_{ijk,0})^2}{l_{ijk,0}}    \label{eq:H_fiber}
\end{equation}
where the parameter, $\mu_m$ is the stretching modulus of the matrix, $l_{ij,0}$ is the bond length prior to any deformation of any bond between two nodes, $l_{ij}$ is the current bond length, $\mu_f$ is the stretching modulus of the fiber, $\kappa_f$ is the bending rigidity of the fiber, $\theta_{ijk,0}$ is the initial angle between the bonds $ij$ and $jk$, $\theta_{ijk}$ is the current bond angle between bonds, $l_{ijk,0} = \frac{l_{ij,0}+l_{jk,0}}{2}$ is an average of the rest length of the two adjacent bonds under consideration, $A_{h,0}$ is the initial area of the hexagon over which the area constraint is being applied and $A_h$ is the current area of the honeycomb. $\beta$ determines the strength of the applied local volume constraint or the strength of resistance to compression and is sometimes also referred to as volume elasticity in confluent models \cite{merkel2019minimal}. Thus the total energy of the composite system is the sum of the two aforementioned individual components, $H_\mathrm{total} = H_\mathrm{matrix} + H_\mathrm{fiber}$. In order to obtain sufficient statistics, about 50 different realizations were analyzed, and the ensemble average of the parameters was computed across these samples. 

To simulate the rheology of the composite, we apply quasi-static shear strain, and then find a mechanically stable equilibrium configuration of the system. The system is periodic in both directions and we use Lees-Edwards boundary conditions to apply shear strain \cite{lees1972computer}. For each strain step, $\delta \gamma$, we first affinely deform the node positions and then obtain the minimum energy configuration of the system using FIRE \cite{bitzek2006structural}. The algorithm is designed to stop when the maximum force on the nodes reaches a value less than a tolerance value (we choose to be $10^{-10}$), which serves as the stopping criteria. The nonaffine node positions are determined as a result of this structural relaxation after which the required macroscopic quantities such as the stress tensor are calculated as \cite{Shivers2019},
\begin{equation}
 \sigma_{ab} = \frac{1}{2V} \sum_{\langle ij \rangle} f_{ij}^a (u_{i}^b - u_{j}^b)    \label{eq:virialstress}
\end{equation}
where, $a$ and $b$ represent any two coordinates, $i$ and $j$ are the nodes connected by a bond, $f_{ij}^a$ is the $a$ component of the force exerted on node $j$ by node $i$, and $u_i^b$ is the $b$ component of displacement vector of node $i$. The stress calculated in equation \ref{eq:virialstress} quantifies the resistance to stretching, bending and compressibility in different directions as a result of the applied external deformation and the enforced energy constraints.
To capture the nonlinearity in the mechanics and to compare with the rheology of experimental systems, we compute the differential shear modulus, $K$, defined as,
\begin{equation}
 K = \frac{d\sigma_{\parallel}}{d\gamma}    \label{eq:K}
\end{equation}
where, $\sigma_{\parallel}$ and $\gamma$ are the shear stress and the applied strain, respectively. To explore the physical implications of $\beta$ and to quantify its effect on the bulk modulus of the matrix and composite, we measure the normal stresses at an applied bulk strain, $\epsilon$. Similar to shear strain, at every step, we apply a small isotropic bulk strain to the composite network and subsequently relax the system. We compute the bulk modulus as a derivative of the normal stress, $\sigma_{\perp}$ i.e. $B = \frac{1}{2}(1+\epsilon)\frac{d\sigma_{\perp}}{d\epsilon}$ \cite{arzash2022mechanics}.

The parameter values are chosen to replicate collagen-HA in physical settings. In a previous study of physical systems, the ratio $\kappa/\mu_f$ for fibers was found to be directly proportional to the fiber volume fraction, $\phi$ \cite{Licup2015}. So, to compare with experiments, values of $\kappa/\mu_f$ less than $10^{-3}$ are suitable, and for most purposes in this study, we set $\mu_f = 1.0$ and $\kappa = 10^{-4}$. The matrix is a soft elastic network with a low stretch modulus of $\mu_m = 10^{-5}$ relative to the the fiber stretch modulus. To evaluate the effect of local incompressibility, $\beta$ is varied to analyze the system up to the physical limits discussed later in this text.

\section{\label{sec:results} Results}

\begin{figure}[h]
\centering
\includegraphics[width=0.45\textwidth]{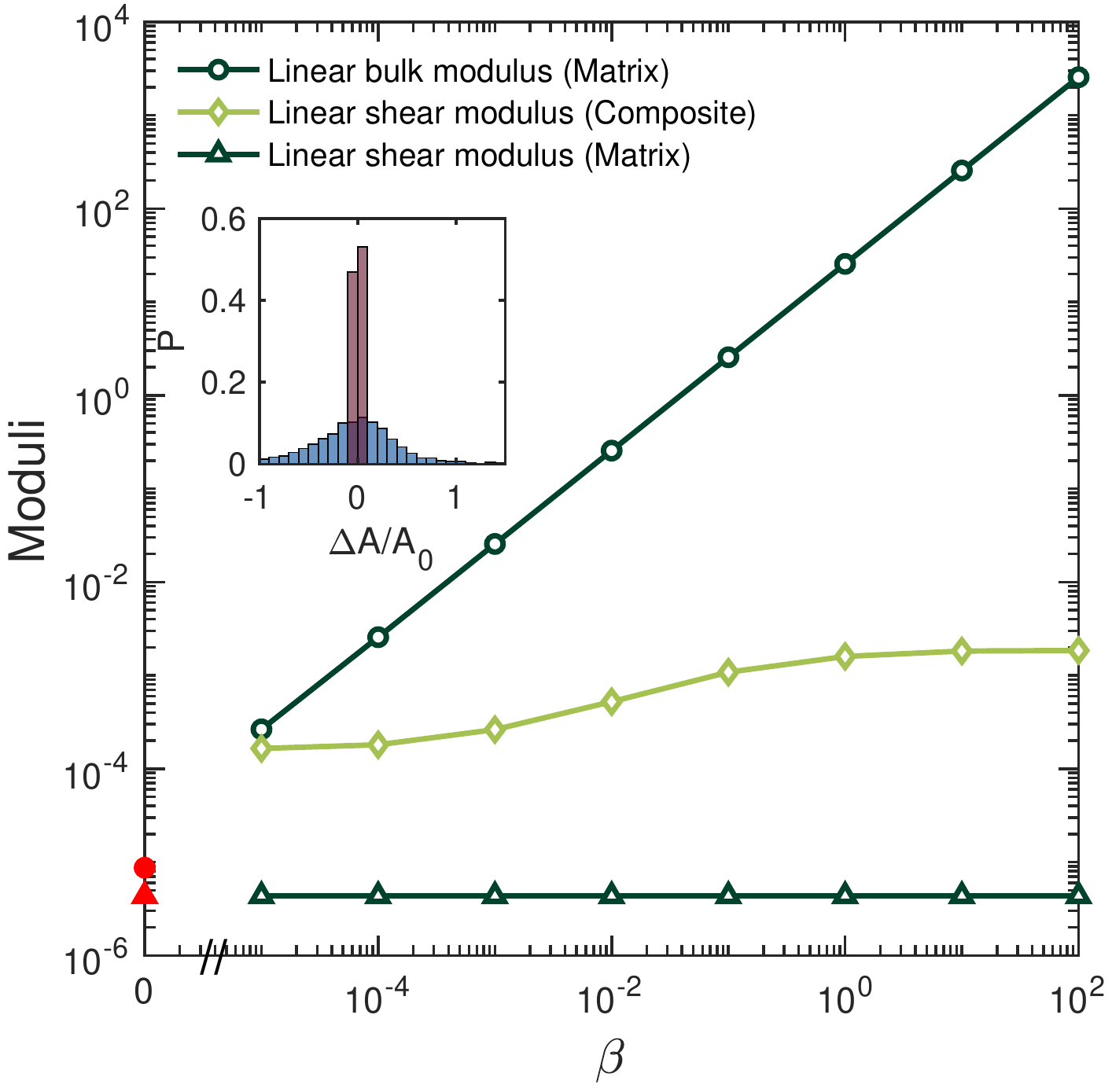} 
\caption{Comparison of shear and bulk moduli of the matrix and composite with varying strength of incompressibility. The shear modulus of the matrix remains unchanged with different values of $\beta$, while the bulk moduli varies proportionally to $\beta$. Therefore, the two moduli of the network are decoupled and can be changed independent of each other and the bulk modulus can be tuned only for a nominal change in the linear shear modulus. (Inset) Area distribution (normalized, P) of the hexagonal regions in the composite on which a local volume constraint has been imposed. When a high level of local incompressibility (shown in red) is applied, the resulting distribution is significantly narrower compared to the distribution without any local incompressibility constraint (shown in blue). The red circle and the triangle represent the values of bulk modulus and shear modulus of the matrix at $\beta = 0$.}
\label{fig:modulus}
\end{figure}

Figure\ \ref{fig:modulus} illustrates the ability of the second term in Eq.\ \eqref{eq:H_matrix} 
to independently control the shear modulus and bulk modulus of both the matrix and composite networks. Previous studies have achieved such a decoupling by tuning the network's microscopic structure \cite{goodrich2015principle, Schlegel2016}. Here, the area constraint provides a simple yet robust decoupling of the elastic moduli for a perfect lattice representing the matrix and using only Hookean terms. 
By increasing the area rigidity $\beta$, the shear modulus of the matrix is unchanged, while the bulk modulus increases linearly with $\beta$. This is consistent with an affine deformation as expected for the full lattice with connectivity above the isostatic threshold. It is noteworthy that the shear modulus of the composite, combining $H_\mathrm{matrix}$ and $H_\mathrm{fiber}$, is affected by the area rigidity $\beta$, suggesting that the resulting deformation is not simply affine. 

To confirm the presence of a local volume constraint, we plot the area distribution of hexagons in the inset to Fig.\ \ref{fig:modulus}. This distribution is shown for two extreme cases of vanishing and large area rigidity $\beta$, corresponding to compressible and incompressible limits. As expected for a local area constraint, the area distributions of hexagonal units are tightly distributed around their initial values for large $\beta$. It is noteworthy that local areas are well-preserved even when the system is subjected to substantial shear strains.

\begin{figure}[h]
\includegraphics[width=0.45\textwidth]{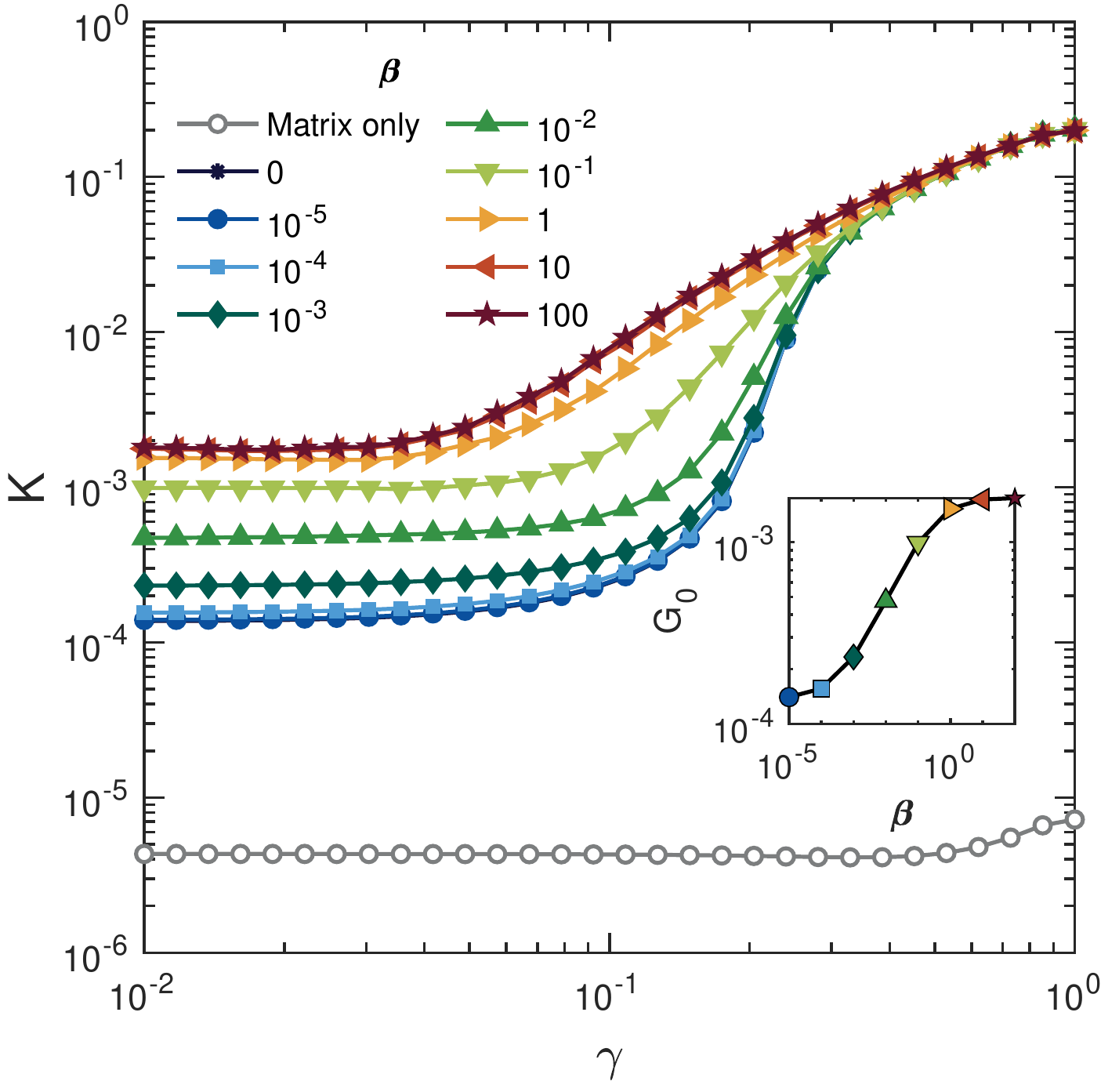}
\caption{ Effect of local incompressibility on the rheology of the composite network. The differential modulus $K$ versus applied shear strain $\gamma$ for different values of area rigidity, $\beta$ on the matrix alone (open symbols) and the composite (closed symbols). (Inset) Linear shear modulus, $G_0$ as a function of $\beta$. Parameter values: $\mu_f = 1.0$, $\mu_m = 10^{-5}$, $\kappa = 10^{-4}$.}
\label{fig:K}
\end{figure}
\begin{figure}[h]
\includegraphics[width=0.45\textwidth]{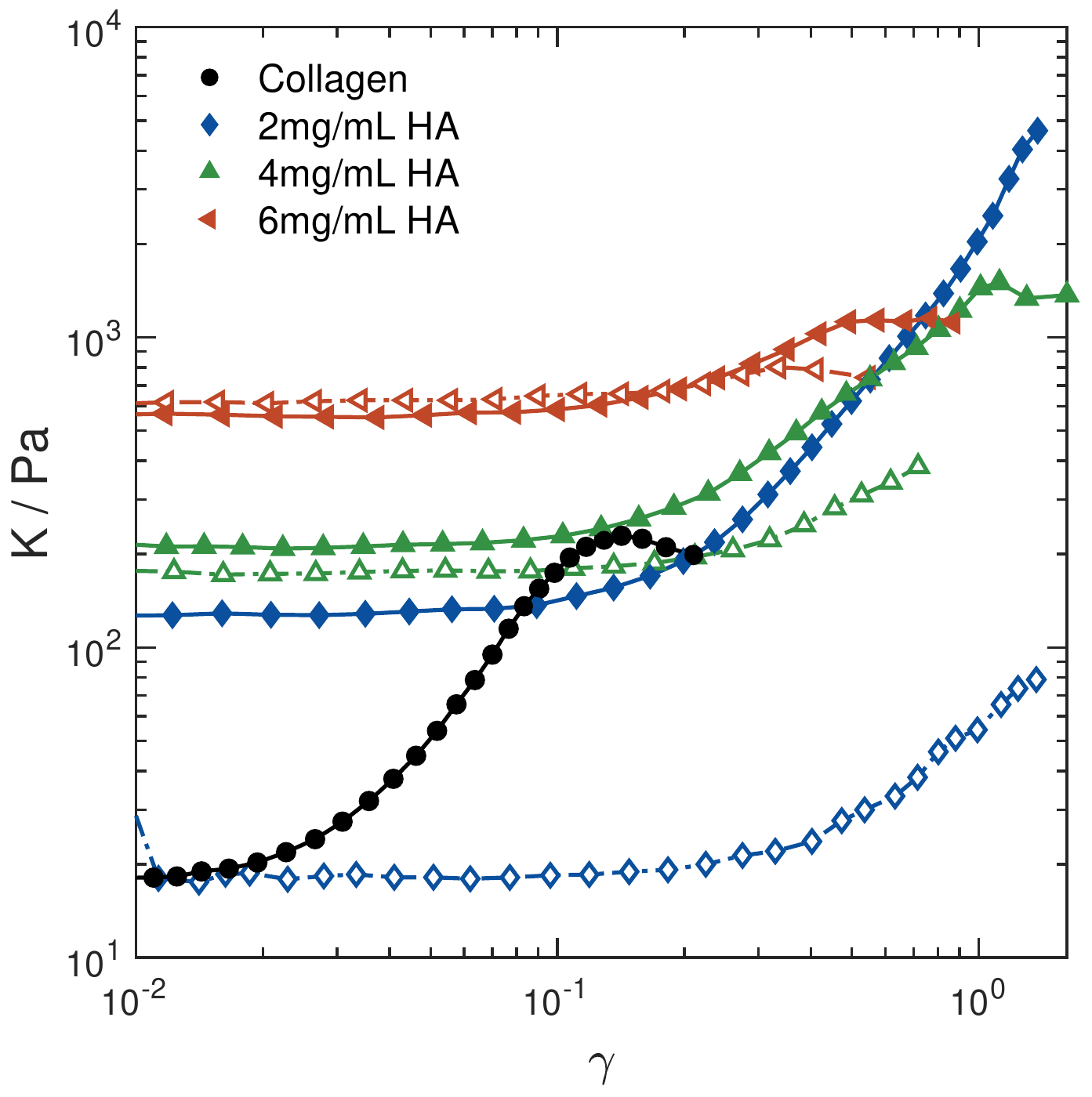}
\caption{Experimental results measuring for the shear of collagen, HA and their composite. The differential modulus, $K$, versus the shear strain for different compositions. Solid lines show pure collagen (black) and composites. Dotted lines show pure HA samples. The differential modulus converges at high strain in each of the composite samples and increases with increasing HA concentration at low strain. The linear modulus of the composite is much higher than the sum of the linear moduli of individual components for a low concentration of HA (2mg/mL). At higher concentrations of HA, the linear modulus of HA dominates, resulting in comparable linear moduli of the composite and pure HA system.}
\label{fig:experiment}
\end{figure}
\begin{figure}[h]
\includegraphics[width=0.475\textwidth]{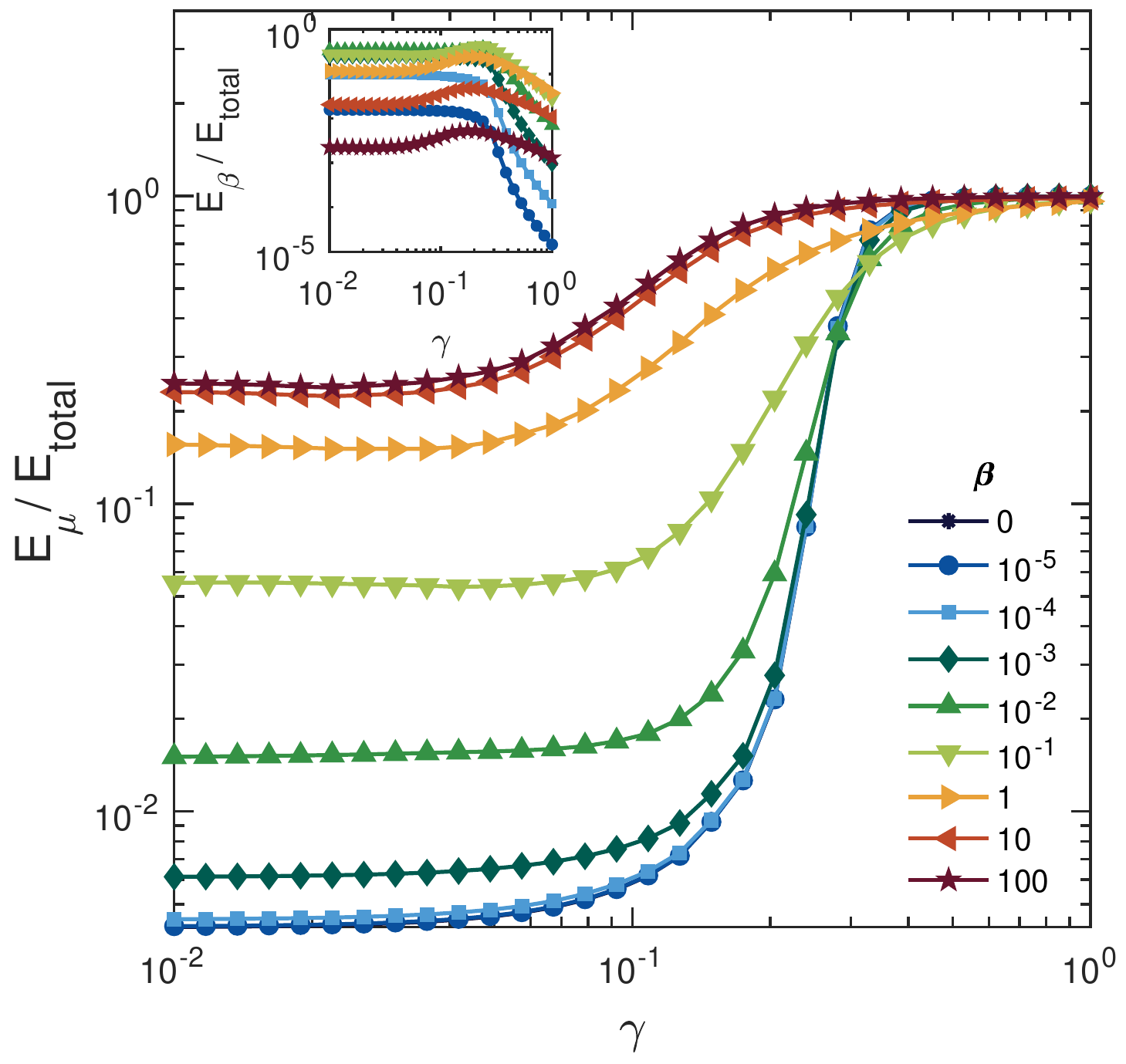}
\caption{Energy contributions to the total energy due to incompressibility, $E_\mathrm{total}$ (inset) and stretching of the fibers, $E_{\mu}$ with varying $\beta$. For $\beta \gtrsim 10^{-2}$, the stretching of fibers is in tandem with values of nonaffinity.}
\label{fig:energies}
\end{figure}

The addition of external or internal energy via bending, prestress, or temperature fluctuations can stabilize subisostatic fiber networks that would otherwise be floppy \cite{sharma2016strain, arzash2019stress, broedersz2011molecular, Dennison2013}. Likewise, the inclusion of area constraints produces a similar stabilizing effect, leading to a non-zero shear modulus within the linear regime. However, as can be seen in Fig.\ \ref{fig:K}, which shows stiffness as a function of stain, the effect of the area rigidity $\beta$ on the resulting shear modulus of the composite is limited, specifically with moduli that remain below the affine limit ($K$ of order unity here). This suggests that under shear, the composite network is able to deform nonaffinely to reduce stretching, even in the presence of the volume constraint. 
For vanishing $\beta$, the fact that the composite shear modulus greatly exceeds the shear modulus for the matrix alone indicates that the elasticity is dominated by the fiber (i.e., $H_\mathrm{fiber}$ above). Moreover, the fact that the linear shear modulus for the composite lies well below the stretching-dominated stiffness at high strain indicates that the linear elasticity is bend-dominated, consistent with prior simulations of such fiber networks \cite{onck2005alternative,sharma2016strain,Broedersz2014} that have shown this to be due to significant nonaffine deformations of the network. 
Interestingly, upon inclusion of the area rigidity $\beta>0$, the linear modulus increases, eventually saturating for large $\beta$ at a value still well below that expected for a purely affine deformation. Thus, although the area constraint appears to suppress some of the nonaffine bending modes, substantial nonaffine deformation is still possible for locally volume preserving networks. 

The nonaffine movements of rigid fibers within the soft matrix give rise to a collective synergistic effect, wherein the linear shear modulus of the composite exceeds the sum of the moduli of the individual fiber and matrix networks (inset of Fig.\ \ref{fig:K}). This effect has been reported in prior experiments on composite biopolymers as well as several synthetic double networks \cite{Doorn2017, Rombouts2013, burla2019stress, rombouts2014synergistic, wyse2022structural}.  Several coarse-grained models have considered the solvent as free-draining within a fibrous network \cite{henriporosity, head2019nonaffinity}, which can also lead to an effective volume constraint on short timescales due to the viscous coupling of solvent and network, together with the incompressibility of the solvent. The timescale for this is determined by the solvent viscosity and network pore size, as discussed above. 
In the case of HA, however, since it forms an entangled meshwork with a much smaller pore size than collagen, it is expected to remain incompressible on longer timescales. Moreover, since the HA and collagen networks are topologically entangled, the resulting volume constraint for the collagen can persist for long times. 
The resulting effect of HA on composites with collagen therefore involves an interplay between the nonaffine bending modes of the collagen network with the local volume constraining effect of the HA. 

To compare our simulations with experiment, we conducted stress ramp measurements on interpenetrating networks of collagen and HA (for details see appendix). Consistent with our simulations, we observed a substantial increase in the linear elastic modulus when collagen and HA were combined, surpassing the individual components' moduli (see Fig.\ \ref{fig:experiment}). This enhancement was particularly prominent at lower concentrations of HA, whereas the linear mechanics of HA predominated the composite network stiffness at higher concentrations. Estimating the value of $\beta$ in experimental setups presents challenges, as it is difficult to isolate the shear modulus of the HA matrix from the incompressibility effect. Our simulations, however, facilitated the disentanglement of this effect, shedding light on the contributions of each component. Additionally, at high strains, the differential moduli converge, aligning with the high strain behavior observed in the simulations (see Fig.\ \ref{fig:K}). Notably, the composite systems exhibited significantly higher strain tolerance to fracture compared to the pure collagen network. Furthermore, the onset strain for nonlinearity was much larger in the collagen-HA composites compared to pure collagen networks. We speculate that collagen polymerization within the HA matrix may lead to reduced collagen connectivity or variations in the thickness of collagen fiber bundles. Nevertheless, it is evident that the linear region clearly demonstrates a synergistic effect consistent with reduced compressibility.

In order to understand the effect of $\beta$ on the rheology of composite networks, it is crucial to analyze the interplay of various energies involved. Figure\ \ref{fig:energies} shows the behavior of the ratio of fiber stretching energy, $E_{\mu}$, to the total energy, $E_\mathrm{total}$, versus strain when varying area rigidity, $\beta$. When $\beta$ is small ($\beta < \kappa$), the composite primarily relaxes through its soft bending modes within the linear regime. Consequently, the stretching modes of the fibers remain relatively inactive, as evidenced in Fig.\ \ref{fig:energies}. However, as $\beta$ increases, the cost associated with the nonaffine bending modes rises since they entail some change in the local density (and, thus, the areas of the hexagonal units in Fig.\ \ref{fig:network}).
As $\beta$ increases, avoidance of local volume (area) change necessitates increasing stretching of the fibers, as is reflected in the increasing ratio of $E_{\mu}/E_\mathrm{total}$. 
However, for large values of $\beta$, the linear modulus appears to saturate to a maximum value (inset of Fig. \ref{fig:K}). This is consistent with a fully incompressible network, for which any further increase in area rigidity becomes irrelevant.  Interestingly, as noted above, the maximal linear modulus still lies below that of a purely affinely deforming network, suggesting the presence of remaining nonaffine deformations that reduce the shear modulus below that for uniform (affine) strain while still preserving the local density of the network. 

\begin{figure}[h]
\includegraphics[width=0.475\textwidth]{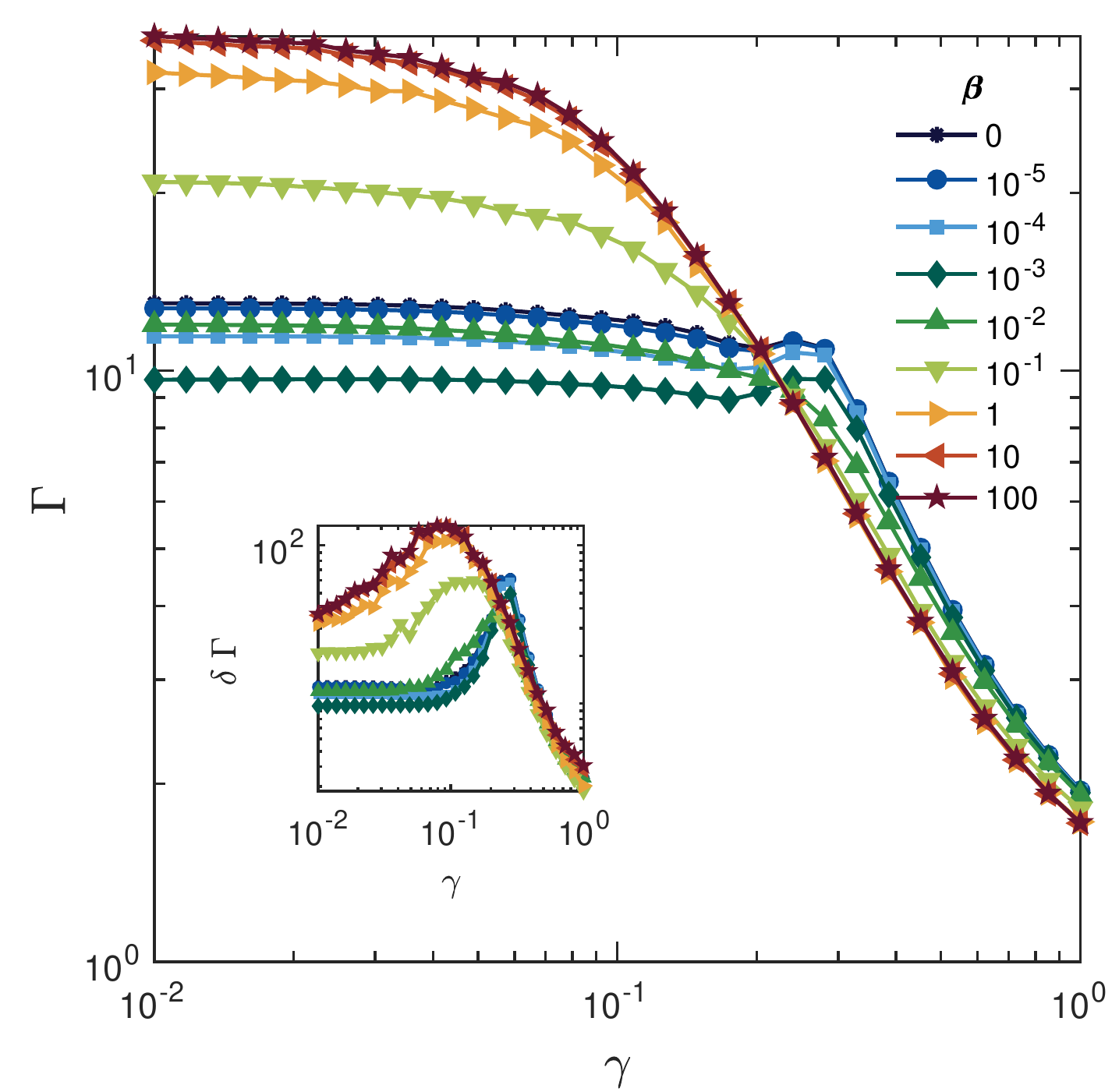}
\caption{ Nonaffinity and differential nonaffinity (inset) of the composite with varying area rigidity $\beta$. The enhancement in the nonaffine fluctuations with $\beta$ in the linear region explains the rise in linear shear modulus (see Fig.\ \ref{fig:K}).}
\label{fig:nonaffinity}
\end{figure}

To gain insights into the interplay of area rigidity with nonaffine deformation, we quantify the nonaffinity fluctuations, given by \cite{Broedersz2011}
\begin{equation}
	\Gamma = \frac{ \langle || \delta \textbf{u}^\mathrm{\textbf{NA}}||^2 \rangle }{l_0^2 \gamma^2 },    \label{eq:nonaffinity}
\end{equation}
as a function of strain. Here,  $\delta \textbf{u}^\mathrm{\textbf{NA}} = \textbf{u} - \textbf{u}^\mathrm{\textbf{affine}}$ represents the difference between the current node displacements and the corresponding affine displacements at the applied strain $\gamma$ and $l_0$ is the average initial bond length, which is $1$ in our model. The brackets denote averaging over all nodes of the network. Figure\ \ref{fig:nonaffinity} shows this nonaffinity versus strain for varying the area rigidity $\beta$. For small values of $\beta\lesssim 10^{-2}$, little change in $\Gamma$ is seen. This is consistent with the prior observation of weak stretching and soft bending modes that account for the nonaffine deformation \cite{Shivers2019, sharma2016straindriven}. The presence of the soft background matrix does not significantly alter this behavior. 
With increasing $\beta\gtrsim 10^{-2}$, the increase in $\Gamma$ is consistent with the increase in stretching noted earlier in Fig.\ \ref{fig:energies}. 

It has previously been observed that the nonaffinity can be identified with critical fluctuations of a second-order-like transition as a function of strain. To analyze this, we consider the differential measure of nonaffine displacements, as introduced in Ref.\ \cite{sharma2016strain}
\begin{equation}
\delta \Gamma = \frac{ \langle || \delta \textbf{u}^\mathrm{\textbf{NA}}||^2 \rangle }{l_0^2 \delta \gamma^2 },    \label{eq:diff_nonaffinity}
\end{equation}
where the nonaffine motions of nodes are measured under an infinitesimal strain step $\delta \gamma$. As shown in the inset of Fig. \ref{fig:nonaffinity}, the criticality of composite networks is apparent in the sharp peak in $\delta \Gamma$ for small values of $\beta\lesssim 10^{-2}$ near the critical strain $\gamma_c\simeq 0.25$. Here, the weak area constraint does not significantly affect the criticality. 
For values of $\beta\gtrsim 10^{-2}$, however, the criticality is strongly suppressed, with a broadening peak in $\delta\Gamma$, spreading to lower values of strain. This is also consistent with the appearance of increasing stretching of fibers in this regime, since the strong critical signatures have been shown to be associated with a sharp transition from bending- to stretching-dominated regimes \cite{sharma2016strain, rens2016nonlinear}. Again, this behavior saturates for large values of $\beta$, consistent with a fully incompressible limit that is insensitive to the value of the area rigidity. 

\section{\label{sec:conclusion} Conclusion}

Here, we have shown theoretically how a local volume constraint can lead to a synergistic enhancement of the linear shear modulus in composites of fibers and flexible polymers, with a composite modulus of approximately ten times that of the sum of the individual moduli for the two components alone. Our model is supported by experiments on collagen-hyaluronan (HA) composites and can also explain other recent experiments on similar composites \cite{Doorn2017, Rombouts2013, burla2019stress, rombouts2014synergistic, wyse2022structural}. 
While we find good qualitative agreement between theory and experiment for the linear modulus, more work will be needed to compare the full nonlinear elasticity. So far, we see a clear suppression of the main feature associated with mechanical criticality, namely the rapid increase in $K$ with strain for the pure collagen sample in Fig.\ \ref{fig:experiment}, consistent with a strong local volume constraint (large $\beta$) in our model. However, the apparent softening of the composite response relative to the pure collagen sample above $~10\%$ strain is not expected within our model. In future work, it will be interesting to vary the HA concentration more systematically at and below 2 mg/mL. It may also be interesting to reduce the HA molecular weight to weaken the effect of topological entanglement with collagen, thereby reducing the volume constraint for the collagen network. 

Theoretically, the addition of the volume constraint with corresponding rigidity $\beta$ is distinct from prior work showing that additional interactions, such as springs or bending rigidity of fibers, can lead to a purely affine response \cite{sharma2016straindriven, Doorn2017, head2003deformation, wilhelm2003elasticity, broedersz2012filament}. Here, the additional rigidity to volume change can stiffen the composite gel, but not to the extent of a purely affine response. As we have shown, this corresponds to remaining non-affine fluctuations that are possible even in the presence of a strong volume constraint: i.e., even volume-preserving non-affine deformations can soften the response of a gel. This is also an area worthy of future study. It may be interesting and fruitful to decompose the non-affine displacement field into divergence-free or volume-preserving and curl-free or potential contributions. 

Finally, the model represented by Eq.\ \eqref{eq:H_matrix} can provide a computationally efficient way to model systems with independent control of bulk and shear moduli, using purely Hookean contributions in the Hamiltonian. Ordinarily, any structure formed with simple springs will lead to a Poisson ratio $\nu=1/4$ in 3D, corresponding to a fixed ratio of bulk to shear moduli of $B/G=5/3$ in 3D \cite{greaves2013poisson, poisson1827note}. (In 2D, $B/G=2$ for spring networks.) In fact, for any system with purely central-force interactions, $\nu=1/4$ in 3D. Such systems are known as Cauchy solids. Thus, with purely central-force interactions, it is not possible to model incompressible materials, for which $B\gg G$ and $\nu=1/2$ in 3D or $\nu=1$ in 2D. 

Although specific values of $B$, $G$ and $\nu$ are not necessarily well-defined for composite materials, even isotropic ones, HA hydrogels should be treated as incompressible on the timescale of experiments such as those presented here. This incompressibility of the gel arises from the incompressibility of the aqueous solvent, together with the viscous coupling of solvent and polymer, which becomes very strong on experimentally-relevant timescales due to the small pore size of HA. For composites of HA plus collagen, as we have discussed above, the incompressibility of the collagen network depends also on the topological entanglement of HA with collagen. In principle, this makes the collagen meshwork incompressible even for very large pores formed by the collagen fibers. The model proposed here can be the basis for efficient future simulations of extracellular matrix composites in this limit. 

\section{\label{sec:acknowledgement} Acknowledgements}

This work is supported by National Science Foundation Division of Materials Research (Grant no. DMR-1826623, DMR-2224030) and the National Science Foundation Center for Theoretical Biological Physics (Grant No. PHY-2019745).  

We thank Matteo d'Este for the kind gift of Tyramine-modified HA and M. Fenu and G. van Osch for useful discussions on the experiments. IM and GHK acknowledge funding from the Convergence programme Syn-Cells for Health(care) under the theme of Health and Technology.



\section{\label{sec:appendix} Appendix}
\begin{center}
    \textbf{Materials and methods}
\end{center}

Experimental rheology tests were performed on Type I bovine (atelo)collagen (Advanced BioMatrix, FibriCol(\textregistered) solution, stock concentration 10 mg/ml in 0.1N hydrochloric acid, Lot 8393) and Tyramine modified hyaluronic acid (Tyr-HA; provided by Matteo D’Este, AO Research Institute Davos, MW 250 kDa, degree of substitution 11\%, lyophilized). Tyr-HA was hydrated in PBS (Sigma Aldrich) with a final concentration of 0.2 U/ml horseradish peroxidase (HRP; Sigma Aldrich, stored at -20\textdegree C and thawed before use). The final Tyr-HA solution concentration was 2 mg/ml, 4 mg/ml or 8 mg/ml. The Tyr-HA solution was left overnight, rotating in a tabletop revolver to dissolve the Tyr-HA in PBS to fully hydrate at 4\textdegree C. On ice, collagen was added to the Tyr-HA solution to a final concentration of 1.5 mg/ml. The composition of the final buffer solution was PBS (140 mmol/l NaCl, 10 mmol/l phosphate buffer, 3 mmol/l KCl), 0.2 U/ml HRP and was set to pH 7.3 by the dropwise addition of 0.1 M NaOH. Finally, hydrogen peroxide (Sigma Aldrich) was added to initiate the cross-linking of the Tyr-HA at a final concentration of 1.5 mmol/l. The solution was mixed by pipetting the solution up and down 10 times then transferring to the bottom plate of the rheometer. This mixing had to be performed quickly so that the sample did not gel in the pipette tip.
For the rheology we used an Anton Paar Physica MCR 501 rheometer (Anton Paar, Graz, Austria) with a stainless steel cone-plate geometry (30 mm diameter, 1 degree angle). The sample was surrounded by mineral oil to prevent solvent evaporation. The temperature was set to 37\textdegree C using a thermostatic hood and a Peltier element (H-PTD200, Anton Paar, Graz, Austria). After loading the sample, the sample-air interface was sealed with heavy mineral oil (Sigma Aldrich, lot \# MKCK4437) and the sample was allowed to polymerize for 1 hour for pure Tyr-HA samples and 2 hours for samples containing collagen. A shear stress ramp from 0.01 Pa to 20 kPa with 10 points per decade, logarithmically spaced, was applied to measure the non-linear rheology of each sample. The measured shear stress and shear strain were converted to differential elastic moduli by taking a numerical gradient of shear stress with respect to shear strain using numpy.


\bibliography{bibliography}

\end{document}